\documentclass[pra,twocolumn,showpacs,groupedaddress,superscriptaddress,aps,10pt]{revtex4-1}
\usepackage{bm,graphicx,amsmath}
\usepackage{placeins}
\usepackage{amssymb}
\usepackage{amsmath}
\usepackage{epsfig}
\usepackage{amssymb}
\usepackage{color}
\usepackage[colorlinks,linkcolor=blue,citecolor=blue]{hyperref}
\usepackage{subfigure}

\begin{document}

\title{Generating a three dimensional dark focus \\ from a single conically refracted light beam}
\date{\today}

\author{Yu. V. Loiko}
\affiliation{Departament de F\'isica, Universitat Aut\`onoma de Barcelona, Bellaterra, E-08193, Spain}
\affiliation{Photonics and Nanoscience Group, School of Engineering, Physics and Mathematics, University of Dundee, Dundee DD1 4HN, UK}
\author{A. Turpin}
\affiliation{Departament de F\'isica, Universitat Aut\`onoma de Barcelona, Bellaterra, E-08193, Spain}
\author{T. K. Kalkandjiev}
\affiliation{Departament de F\'isica, Universitat Aut\`onoma de Barcelona, Bellaterra, E-08193, Spain}
\affiliation{Conerefringent Optics SL, Avda. Cubelles 28, Vilanova i la Geltr\'u, E-08800, Spain}
\author{E. U. Rafailov}
\affiliation{Photonics and Nanoscience Group, School of Engineering, Physics and Mathematics, University of Dundee, Dundee DD1 4HN, UK}
\author{J. Mompart}
\affiliation{Departament de F\'isica, Universitat Aut\`onoma de Barcelona, Bellaterra, E-08193, Spain}

\begin{abstract} 
We report here the generation of a three dimensional dark focus from a single focused monochromatic Gaussian beam that undergoes conical refraction when it propagates along one of the optic axes of a biaxial crystal. We study the resulting ring intensity pattern behind the crystal as a function of the ratio between the ring radius and the beam waist and derive the particular parameter values for which a three dimensional dark focus with null intensity at the ring center is formed. We have performed experiments with a ${\rm KGd(WO_4)_2}$ biaxial crystal reporting the generation of a bottle beam in full agreement with our theoretical investigations.  \\
\textbf{ocis}: 260.1440, 020.7010, 020.3320, 350.3390
\end{abstract}

\date{\today}
\maketitle

Optical beams with dark regions of exactly zero intensity are rare objects usually related to optical singularities. Their applications range from particle trapping \cite{2010_Shvedov_PRL_105_118103}, subdiffraction limited tighter focusing \cite{2004_Sheppard_AO_43_4322} and plasmon excitation \cite{2007_Bouhelier_OL_32_2535} to laser machining \cite{2007_Meier_APA_86_329}. 
Laguerre--Gauss beams are well known examples of light beams possessing optical vortices forming a straight nodal line surrounded by light \cite{1992_Allen_PRA_45_8185,Blio2010}. More involved structures of closed loop nodal lines, their threading, knotting and linking have been demonstrated recently \cite{2001_Berry_PRSA_457_2251, 2001_Berry_JPA_34_8877, 2004_Leach_Nature_432_165, 2008_Irvine_NatPhys_4_716, 2010_Dennis_NatPhys_6_118}.

Bottle beams \cite{2000_Arlt_OL_25_191, 1995_Davidson_PRL_74_1311, 2001_Cacciapouti_EPJD_14_373, 2001_Rudy_OE_8_160, 1997_Kuga_PRL_78_4713, 2012_Li_OL_37_851, 1999_Ozeri_PRA_59_R1750, 2004_Yelin_OL_29_661, 2009_Isenhower_OL_34_1159} are another example of optical beams with a dark region, more precisely they are also named optical dark potentials. Ideally, bottle beams comprise zero electric amplitude at one spatial point surrounded in all directions by regions with relatively high intensity. In practice, the intensity minimum in bottle beams is not exactly equal to zero because of different experimental imperfections. Various methods and techniques have been proposed to produce three dimensional (3D) optical dark potentials in a controllable way, such as creating an intensity minimum by surrounding a region in 3D space with several beams \cite{1995_Davidson_PRL_74_1311, 2001_Cacciapouti_EPJD_14_373, 2001_Rudy_OE_8_160}, crossing at least two cylindrical-vector and vortex beams with a phase dislocation along the beam axis that leads to a zero-intensity point \cite{1997_Kuga_PRL_78_4713, 2012_Li_OL_37_851}, by destructive interference of several Laguerre--Gauss light beams \cite{1999_Ozeri_PRA_59_R1750, 2000_Arlt_OL_25_191,2004_Yelin_OL_29_661,2009_Isenhower_OL_34_1159} or using uniaxial c-cut crystals \cite{kro_OE_2008, kro_JOSAB_2013}. However, these methods have several drawbacks such as the fact that in some of them the intensity minimum is not exactly equal to zero, the extreme precise control on the optical elements being used, the field fluctuations introduced at and close to the zero-amplitude point, or the non trivial generation of Laguerre-Gauss and cylindrical-vector and vortex beams \cite{2009_Zhan_AOP_1_1}. 
Optical bottle beams have applications in optical tweezers for trapping particles with a refractive index lower than the surrounding medium, using the photophoretic force for instance \cite{2010_Shvedov_PRL_105_118103}, or in cold atom trapping with the possibility of creating an all optical blue-detuned trap that operates in the zero-intensity region \cite{2001_Bongs_PRA_63_031602}. 

In this Letter, we present a robust, easy, and compact alternative to the previously cited methods to generate an optical bottle beam by transforming a fundamental monochromatic Gaussian beam into a beam with a 3D dark focus, i.e. a beam with a point of exact null intensity surrounded in all directions by regions of higher intensity, using the conical refraction (CR) phenomenon occurring in biaxial crystals \cite{1978_Belskii_OS_44_436,1999_Belsky_OC_167_1,2004_Berry_JOA_6_289,2007_Berry_PO_55_13,1942_Raman_Nature_149_552,2008_Kalkandjiev_SPIE_6994,2013_Sokolovskii_OE_21_11125,mcd2012,CRlaser2010a,CRlaser2010b}.

Let us consider a focused monochromatic input beam propagating along one of the optic axis of a biaxial crystal. In what follows, we will assume a fundamental Gaussian input beam whose Rayleigh length and waist radius are given by $z_R$ and $w_0$, respectively. If focusing lenses are used, the beam waist position is located at the focal plane of the lens. The crystal is characterized by its length, $l$, and conicity parameter, $\alpha$, that is half the intersection angle of the two ellipsoids of refractive indices \cite{1978_Belskii_OS_44_436, 1999_Belsky_OC_167_1, 2004_Berry_JOA_6_289}. Their product, $R_0=\alpha l$, gives the geometric optic approximation for the CR ring radius behind the crystal. As reported below, the resulting CR intensity pattern depends crucially on the parameter $\rho_0 \equiv R_0 / w_0$. Note that for a fixed $\alpha$, $\rho_{0}$ can be controlled by varying either the crystal length, $l$, or the focused beam waist, $w_{0}$. Here we will vary the latter one modifying the focal length of the lens used in the experiments.

CR is usually associated with the appearance of a bright ring intensity pattern with fine Poggendorff splitting at the focal plane, when a focused light beam passes along one of the optic axes of a biaxial crystal. Note that the position where the CR light ring is formed does not depend on the position of the biaxial crystal with respect to the beam waist plane.
Belsky and Khapalyuk \cite{1978_Belskii_OS_44_436} were the first to derive a paraxial solution for CR from Maxwell equations, this solution later on beingreformulated by Berry  \cite{2004_Berry_JOA_6_289}. 
For a uniformly polarized and cylindrically symmetric input beam, the electric field amplitude behind the crystal can be written as follows \cite{1978_Belskii_OS_44_436}:
\begin{eqnarray}
&&
\mathbf{E} \left( \rho,Z\right)  =
\left(
\begin{array}{cc}
B_{C} + C & S \\ 
S & B_{C} - C
\end{array}
\right)
\mathbf{e}^{\rm{in}}
,~ \label{Eqs_output_beam_uniform}
\end{eqnarray}
where $C=B_{S} \cos \left( \varphi + \varphi_{C} \right)$ and $S=B_{S} \sin \left( \varphi + \varphi_{C} \right)$.
$\mathbf{E}=\left( E_{x}^{\rm{out}}, E_{y}^{\rm{out}} \right)$ denotes the transverse Cartesian components of the output field,
$\mathbf{e}^{\rm{in}}= \left( e_{x}^{\rm{in}}, e_{y}^{\rm{in}} \right)$ is the unit vector of the input electric field, and $\varphi _{C}$ is the polar angle that defines the orientation of the plane  of the crystal optic axes. $Z=z/z_{R}$ and $\mathbf{\rho} = \left( \cos \varphi,\sin \varphi \right) r /w_{0}$ define cylindrical coordinates whose origin is associated to the geometrical center of the light pattern at the focal plane ($Z=0$). $B_{C}$ and $B_{S}$ are the main integrals of the Belsky--Khapalyuk paraxial solution for CR:
\begin{eqnarray}
B_{C}\left( \rho,Z \right) = \int_{0}^{\infty }  \eta a \left( \eta \right)
e^{-i \frac{Z}{4} \eta ^{2} } \cos \left( \eta \rho_{0}\right) J_{0}\left( \eta \rho \right) d \eta
, \label{Eqs_Bc_general} \\
B_{S}\left( \rho,Z \right) = \int_{0}^{\infty }  \eta a \left( \eta \right)
e^{-i \frac{Z}{4} \eta ^{2} } \sin \left( \eta \rho_{0}\right) J_{1}\left( \eta \rho \right) d \eta
, \label{Eqs_Bs_general} \\
a \left( \eta \right) = \int_{0}^{\infty} \rho E^{\rm{in}} \left( \rho \right) J_{0} \left( \eta \rho \right) d \rho
, \label{Eqs_aP_uniform}
\end{eqnarray}
with $a \left( \eta \right)$ being the radial part of the 2D transverse Fourier transform of an input beam
$\mathbf{E}^{\rm{in}}=E^{\rm{in}} \left( \rho \right) \mathbf{e}^{\rm{in}}$,
$\eta = \left| \mathbf{k}_{\perp} \right| w_{0}$
is the modulus of the transverse wavevector components projected onto the entrance surface of the crystal, and $J_{q}$ denotes the Bessel function of order $q$.

For an input beam of circular polarization (CP) (also for random polarization, RP) and of linear polarization (LP), the intensity distribution behind the crystal becomes, respectively:
\begin{eqnarray}
I_{\rm{CP}} \left( \mathbf{\rho},Z \right) 
= \left| B_{C} \right|^{2} + \left| B_{S} \right|^{2}
,~ \label{Eqs_output_beam_intensity_CP}
\\
I_{\rm{LP}} \left( \mathbf{\rho},Z\right) 
= I_{\rm{CP}} + 2 {\rm Re} \left[ B_{C} B_{S}^{*} \right]
\cos \left( 2 \Phi - \left( \varphi + \varphi_{C} \right) \right)
,~ \label{Eqs_output_beam_intensity_LP}
\end{eqnarray}
where $\Phi$ is the polarization azimuth of the LP input beam with $\mathbf{e}^{\rm{in}}= \left( \cos \Phi , \sin \Phi \right)$.
Eqs.~(\ref{Eqs_output_beam_uniform})-(\ref{Eqs_output_beam_intensity_LP}) show that a radially symmetric intensity pattern of CR is obtained for a CP input beam. Instead, for a LP input beam, a crescent ring intensity pattern appears such that zero intensity is obtained for the ring position that possess orthogonal polarization with respect to the input polarization. In what follows, we will focus on the CP input beam case since our aim is to create a 3D dark focus with radial symmetry along the $Z$ axis.  

From Eq.~(\ref{Eqs_output_beam_intensity_CP}), the on $Z$-axis intensity distribution is defined solely by the function $B_{C} \left( \rho=0, Z \right)$ since $J_1(0)=0$ and, therefore, $B_{S} \left( \rho=0, Z \right)=0$, see  Eq.~(\ref{Eqs_Bs_general}). The former can be evaluated analytically for a fundamental Gaussian beam \cite{2004_Berry_JOA_6_289},
$a \left( \eta \right) = \exp \left( -\eta ^{2} / 4 \right)$, yielding:
\begin{eqnarray}
B_{C}( 0, Z) &=& \frac{1}{w_{Z}} \frac{d}{dX} {\rm F} (X) \nonumber \label{Eqs_Gauss_ONaxis_B0_Loiko} \\
&=& \frac{1}{w_{Z}} \left[ 1 + i \sqrt{\pi} X e^{-X^{2}} {\rm erf} ( i X) \right], \label{Eqs_Gauss_ONaxis_B0_Berry} 
\end{eqnarray}
where $X = \rho_{0} / \sqrt{w_{Z}}$ and $w_{Z} = 1 + i Z$. 
${\rm F} (X)$ and ${\rm erf} (X)$ denote Dawson and error functions, respectively.
Dawson function satisfies the equation $d {\rm F} (X) / dX = 1 - 2 X {\rm F} (X)$.

\begin{figure}[]
\centering
\includegraphics[width=\columnwidth]{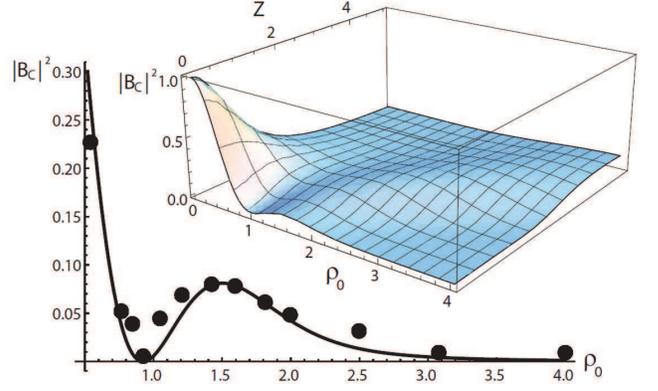}
\caption{(Color online) On-axis intensity at the focal plane center, $I_{Z}=\left| B_{C} \left( \rho=0, Z=0\right) \right|^{2}$, as given by Eq.~(\ref{Eqs_Gauss_ONaxis_B0_Berry}) as a function of the control parameter $\rho_{0}=R_0/w_0$. 
Black solid circles represent the experimental data obtained when a focused Gaussian beam propagates along the optic axis of a ${\rm KGd(WO_4)_2}$ biaxial crystal. The inset shows the on-axis intensity $I_{Z}=\left| B_{C} \left( \rho=0, Z \right) \right|^{2}$. }
\label{figONaxis1}
\end{figure}

For $\rho_{0} \gg 1$, Eqs.~(\ref{Eqs_output_beam_uniform}-\ref{Eqs_output_beam_intensity_LP}) were investigated analytically in Refs.~\cite{1978_Belskii_OS_44_436,1999_Belsky_OC_167_1} at the focal plane, whereas Eq.~(\ref{Eqs_Gauss_ONaxis_B0_Berry}) was studied in Refs.~\cite{2004_Berry_JOA_6_289,2007_Berry_PO_55_13}. In this case, the CR ring with the Poggendorff fine splitting is well resolved at the focal plane. Moreover, two bright spots, called after Raman \cite{1942_Raman_Nature_149_552}, were observed symmetrically located on the $Z$-axis at both sides from the CR ring \cite{2008_Kalkandjiev_SPIE_6994}.

We analyze here the case $\rho_{0} \sim 1$ in 3D paying attention to the transition from a single bright focus in the absence of the crystal to the appearance of a CR ring with well separated Raman spots. Eq.~(\ref{Eqs_Gauss_ONaxis_B0_Berry}) has a solution with zero amplitude at the focus center $\left( \rho=0,Z=0 \right)$ that corresponds to the maximum of the Dawson function ${\rm F} (X)$. Thus, numerically solving $B_{C}( 0, Z)=0$ from Eq.~(\ref{Eqs_Gauss_ONaxis_B0_Berry}), one expects to obtain a dark focus (DF) (zero amplitude) for 
\begin{equation}
\rho_{0}^{\rm{DF}} = 0.924, \label{Eqs_Gauss_X0_DF} \\
\end{equation}
where the superscript `DF' means dark focus.

Fig.~\ref{figONaxis1} shows the evolution of the normalized on-axis intensity $I_{Z}=\left| B_{C} \left( \rho=0, Z =0 \right) \right|^{2}$ at the focal plane as a function of the control parameter $\rho_0 = R_0 / w_0$ (solid line) as given by Eq.~(\ref{Eqs_Gauss_ONaxis_B0_Loiko}), while the inset shows the on-axis normalized intensity distribution $I_{Z}=\left| B_{C} \left( \rho=0, Z \right) \right|^{2}$. The latter clearly shows that a 3D dark focus is expected for $\rho_0 =\rho_0^{\rm{DF}}$. 

%
\begin{figure}[]
\centering
\includegraphics[width=\columnwidth]{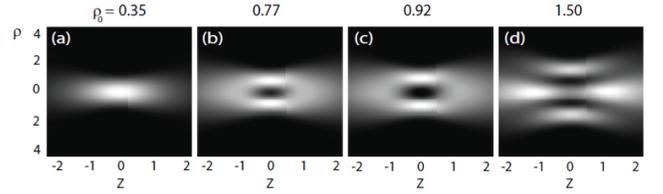}
\caption{ Slice of the intensity distributions $I_{\rm{CP}} \left( \rho, Z \right) $ along the propagation direction for an input Gaussian beam of circular polarization as given by Eqs.(\ref{Eqs_Bc_general})-(\ref{Eqs_output_beam_intensity_CP})
for different values of the control parameter $\rho_{0}$.}
\label{figONcentre1}
\end{figure}

Since the $Z$-axis is the symmetry axis of the output beam, the 3D distribution of the electric field can be represented by a two dimensional slice of intensity $I_{\rho Z} = I \left(\rho, Z \right)$ as shown in Fig.~\ref{figONcentre1}. For $\rho_{0} \leq 0.45$, a single bright spot is observed whose maximum intensity corresponds to the center of the focal plane, see Fig.~\ref{figONcentre1}(a). For larger values of $\rho_{0}$, this maximum is laterally shifted in the focal plane and an
intensity minimum appears at $\rho=Z=0$, see Figs.~\ref{figONcentre1}(b,c). In particular, a 3D dark focus, i.e. a bottle beam, is expected for $\rho_0 =\rho_0^{\rm{DF}}=0.92$, as shown in Fig.~\ref{figONcentre1}(c) that disappears for larger values of $\rho_0$, see Fig.~\ref{figONcentre1}(d). As it is seen in Fig.~\ref{figONcentre1}(c), the 3D dark focus with a zero intensity point at the beam center is surrounded by a region of higher intensity in all directions. The maximum in the intensity barrier is achieved at the focal plane on a ring with radius $\rho_{max} \approx 1.1$ and $I\left( \rho_{max}, Z=0 \right) \approx 0.2$, i.e. the peak intensity at this ring is approximately $20\%$ of that provided by the same focused Gaussian beam without the crystal. Along the beam propagation direction, the axial intensity has maxima at $Z_{max} \approx \pm 1.388$ with $I\left( \rho=0, Z_{max} \right) \approx 0.14$, in good agreement with the expected position for the Raman spots $Z_{\rm{Raman}} \approx \pm \sqrt{2} \rho_{0}$ previously estimated for a well developed CR ring  \cite{2004_Berry_JOA_6_289}.
The minimum in the intensity barrier has a form of a ring with radius $\rho_{\theta}\approx0.62$ that appears at a distance $Z_{\theta}\approx \pm 1.1$ with $\left|\rho_{\theta}/Z_{\theta}\right| \approx \tan 30^{0}$ and $I\left( \rho_{\theta}, Z_{\theta} \right) \approx 0.13$. Dark focus, see Fig.\ref{figONcentre1}(c), and dark ring, see Fig.\ref{figONcentre1}(d), appear because of destructive interference of two cones provided by the CR phenomenon inside biaxial crystal and displaced with respect to each other along propagation direction as detailed in Ref.~\cite{2013_Sokolovskii_OE_21_11125}.

To experimentally confirm the theoretical prediction on the possibility to generate a CR bottle beam with a 3D dark focus, we have performed experiments on conical refraction in a plate of monoclinic ${\rm KGd\left( WO_{4} \right)_{2}}$ single crystal ($\alpha=0.017~\rm{rad}$) cut perpendicular to one of the optic axes. The crystal length $l=2.3~\rm{mm}$ yields a CR ring radius of $R_{0} = 39.1~\rm{\mu m}$. As an input we take a collimated circularly polarized Gaussian beam (with radius of $1.5~\rm{mm}$) obtained from a $635~\rm{nm}$ diode laser coupled to a monomode fiber. The input collimated beam is focused by a lens and the transverse intensity patterns behind the crystal are recorded at different positions along the beam propagation direction within few Rayleigh lengths. By varying the focusing distance of the lens, we are able to adjust the radius of the focused beam, i.e. the parameter $\rho_{0}=R_{0} / w_{0}$ in the range $\rho_{0} \in \left[ 0.3 , 4.0 \right]$.

\begin{figure}[t]
\centering
\includegraphics[width=\columnwidth]{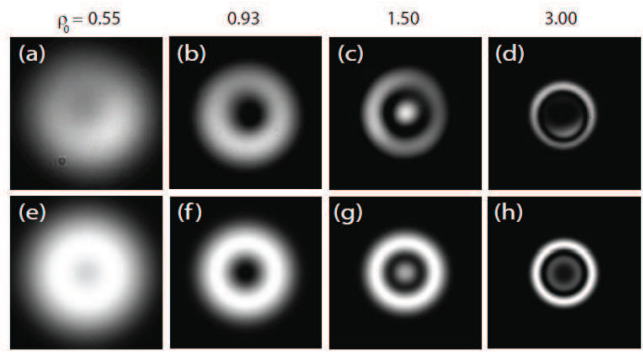}
\caption{ Transverse intensity patterns at the focal plane experimentally obtained (a)-(d) and theoretically calculated (e)-(h), see Eqs.~(\ref{Eqs_Bc_general})-(\ref{Eqs_output_beam_intensity_CP}), for a circularly polarized input beam of fundamental Gaussian profile with different values of the waist radius $w_{0}$, which gives different values of control parameter $\rho_{0}$.}
\label{figW0Denepdence1}
\end{figure}

Normalized intensities at the center of the focal plane experimentally measured are plotted in Fig.~\ref{figONaxis1} with black solid points, while the experimentally captured full transverse intensity patterns are represented in the upper row of Fig.~\ref{figW0Denepdence1} for different values of $\rho_{0}$. As indicated before, for $\rho_0<0.45$, the intensity pattern resembles that of a Gaussian beam. From $\rho_0=0.45$ to $\rho_0=0.92$, the intensity at the center decreases being exactly zero at $\rho_0=0.92$, see Figs.~\ref{figW0Denepdence1}(a) and (b). From $\rho_0=0.92$ to $\rho_0=1.50$, it increases with a relative maximum at $\rho_0=1.50$, see Fig.~\ref{figW0Denepdence1}(c). From this value on, the intensity at the center monotonically decreases and the full transverse intensity pattern becomes the standard pattern of CR with two bright rings, see Fig.~\ref{figW0Denepdence1}(d). The waist of the focused beam has been measured by removing the crystal. For the results shown in Fig.~\ref{figW0Denepdence1}(b), it provides the ratio $\rho_{0}^{\left( \rm{exp} \right)} = 0.93$, which is close to the theoretical value $\rho_{0}^{\rm{DF}}$, see Eq.~(\ref{Eqs_Gauss_X0_DF}).
The lower row in Fig.~\ref{figW0Denepdence1} shows the full transverse intensity patterns theoretically calculated from Eqs.~(\ref{Eqs_output_beam_uniform})-(\ref{Eqs_output_beam_intensity_LP}) being in an almost perfect agreement with the experimental results. 

\begin{figure}[]
\centering
\includegraphics[width=\columnwidth,clip=true]{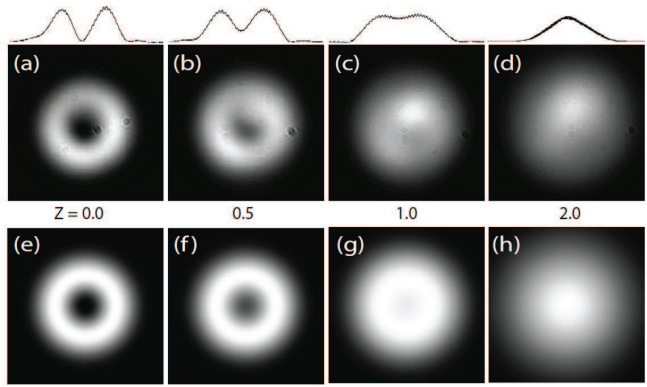}
\caption{Spatial evolution of the transverse intensity patterns along the propagation distance $Z$ for $\rho_{0}^{\left( \rm{exp} \right)} = 0.93 \sim \rho_{0}^{\rm{DF}}$, see Eq.(\ref{Eqs_Gauss_X0_DF}), as experimentally obtained (a)-(d) and theoretically calculated (e)-(h), see Eq.(\ref{Eqs_Bc_general})-Eq.~(\ref{Eqs_output_beam_intensity_CP}). Top insets represent the cross-section of the measured intensity distribution along the horizontal axis at the beam center.}
\label{figLloydPlane1}
\end{figure}

Figs.~\ref{figLloydPlane1}(a)-(d) present the experimentally measured spatial evolution of the transverse intensity patterns for $\rho_{0}^{\left( \rm{exp} \right)} = 0.93 \sim \rho_{0}^{\rm{DF}}$ that corresponds to the case where a 3D dark focus appears at the focal plane. The dark focus pattern from Fig.~\ref{figLloydPlane1}(a) evolves to a pattern with a maximum intensity at its center as shown in Fig.~\ref{figLloydPlane1}(d). Clearly, a region of null intensity is surrounded by regions of higher intensity. The Rayleigh length in this case is $z_{R}\approx 8.9$mm, which gives an estimation of the depth of the dark potential. Figs.~\ref{figLloydPlane1}(e)-(h) are the corresponding theoretical predictions obtained from Eqs.~(\ref{Eqs_output_beam_uniform})-(\ref{Eqs_output_beam_intensity_LP}). Both theory and experiment are in good agreement.

In conclusion, CR has been investigated for the first time both theoretically and experimentally in the regime $\rho_0 = R_0/w_0 \approx 1$ for which novel intensity structures appear. For the specific value $\rho_0=0.924$, we have theoretically derived and experimentally reported the transformation of an input Gaussian beam into a bottle beam with a point of exact null intensity. At variance with spatial light modulators (SLMs), where a significant amount of light is lost by diffraction, the here proposed method transfers all the input power into the bottle beam, being essential for applications such as laser drilling, stimulated emission depletion microscopy, or trapping neutral atoms and Bose-Einstein condensates (BECs) by means of the light dipole force. An additional advantage is the fact that quality and smoothness of the 3D dark focus obtained by means of CR is only limited by the quality of the input beam and the focusing lenses, while for SLMs  it strongly depends on the pixel density. 
Finally, as it is shown in Figs.~3(b) and (c), the bottle beam generated with CR can be adiabatically transformed into a 3D dark ring, by simply tuning the focusing geometry. Hence, it could be possible to trap a BEC in the 3D dark focus and then adiabatically transfer it into the dark ring to investigate, for instance, matter wave (Sagnac) interferometry or the appearance of persistent currents.

The authors gratefully acknowledge financial support through Spanish MICINN contract FIS2011-23719, and the Catalan Government contract SGR2009-00347. This research was partially supported by FP7 Project HiCORE under grant agreement number 314936. A. Turpin acknowledges financial support through grant AP2010-2310 from the MICINN.

\end{document}